\documentclass[aps,prl,twocolumn,superscriptaddress,showpacs,showkeys]{revtex4-1}

\usepackage{graphicx}
\usepackage{siunitx}
\usepackage{amsmath, amsxtra, amssymb, mathtools, isomath, nccmath, xspace, siunitx}
\usepackage{color}
\usepackage[utf8]{inputenc}
\usepackage{hyperref}
\usepackage{sidecap}

\newcommand{\ket}[1]{\ensuremath{\lvert #1 \rangle}\xspace}%
\newcommand{\bra}[1]{\ensuremath{\langle #1 \rvert}\xspace}%
\newcommand{\alat}{\ensuremath{a_{\text{lat}}}\xspace}%

\sidecaptionvpos{figure}{h}

\begin{document}

\title{Microscopy of a scalable superatom} 

\author{Johannes Zeiher}
\email[]{johannes.zeiher@mpq.mpg.de}
\author{Peter Schauß}
\author{Sebastian Hild}
\affiliation{Max-Planck-Institut f\"{u}r Quantenoptik, 85748 Garching, Germany}
\author{Tommaso Macr\`i}
\affiliation{QSTAR, Largo Enrico Fermi 2, 50125  Firenze, Italy}
\author{Immanuel Bloch}%
\affiliation{Max-Planck-Institut f\"{u}r Quantenoptik, 85748 Garching, Germany}
\affiliation{Ludwig-Maximilians-Universit\"{a}t, Fakult\"{a}t f\"{u}r Physik, 80799 M\"{u}nchen, Germany}%
\author{Christian Gross}%
\affiliation{Max-Planck-Institut f\"{u}r Quantenoptik, 85748 Garching, Germany}

\date{\today}

\begin{abstract}
  Strong interactions can amplify quantum effects such that they become important on macroscopic scales.
  Controlling these coherently on a single particle level is essential for the tailored preparation of strongly correlated quantum systems and opens up new prospects for quantum technologies.
  Rydberg atoms offer such strong interactions which lead to extreme nonlinearities in laser coupled atomic ensembles. 
  As a result, multiple excitation of a Micrometer sized cloud can be blocked while the light-matter coupling becomes collectively enhanced.
  The resulting two-level system, often called ``superatom'', is a valuable resource for quantum information, providing a collective Qubit.
  Here we report on the preparation of two orders of magnitude scalable superatoms utilizing the large interaction strength provided by Rydberg atoms combined with precise control of an ensemble of ultracold atoms in an optical lattice.
  The latter is achieved with sub shot noise precision by local manipulation of a two-dimensional Mott insulator. 
  We microscopically confirm the superatom picture by in-situ detection of the Rydberg excitations and observe the characteristic square root scaling of the optical coupling with the number of atoms.  
  Furthermore, we verify the presence of entanglement in the prepared states and demonstrate the coherent manipulation of the superatom. 
  Finally, we investigate the breakdown of the superatom picture when two Rydberg excitations are present in the system, which leads to dephasing and a loss of coherence. 
\end{abstract}



\maketitle

Nonlinearities in light-matter coupling are usually weak, leading to a linear growth of the number of optical excitations with increasing photon flux.
In contrast, the most extreme regime of strong nonlinearities is reached when an ensemble of many absorbers can host only a single excitation, such that one photon already saturates the medium.
This can be realized with the aid of optical cavities~\cite{Munstermann2000} or, in free space, by atomic ensembles excited to Rydberg states~\cite{Saffman2010}.
For the latter, extremely strong dipolar interactions between Rydberg atoms block all but a single optical excitation in a volume of several Micrometers~\cite{Jaksch2000, Lukin2001, Urban2009, Gaetan2009}, effectively transforming the $N$ atoms within this volume to one collective two-level system.
Under uniform illumination this ``superatom'' features enhanced coupling to the light field and the Rydberg excitation is symmetrically shared between the individual atoms~\cite{Dicke1954}.
The resulting singly excited Dicke state is also known as $W$-state whose many-body character is reflected in multipartite entanglement between its constituents~\cite{Haas2014}.
Superatoms are valuable resources for quantum information. 
They have been proposed as collective Qubits~\cite{Lukin2001} and indeed, strong interactions between them were demonstrated recently~\cite{Ebert2015}.
These collective Qubits have distinct advantages over single atoms that have previously been entangled using the strong Rydberg interactions~\cite{Jaksch2000, Isenhower2010, Wilk2010}.
First, the inherent collective enhancement of the atom-light coupling provides a single photon interface and efficient entanglement transfer between atoms and light~\cite{Saffman2005, Pedersen2009, Li2013}. 
Second, the information is redundantly stored in the $N$ constituent particles, protecting it against detrimental atom loss~\cite{Dur2000, Brion2008}.  
Further applications reach from advanced Qubit encoding schemes in multi-level atoms~\cite{Brion2007, Saffman2008} to efficient single photon sources~\cite{Saffman2002, Dudin2012b} and single photon subtraction~\cite{Honer2011a}.
Coherent manipulation of superatoms is at the heart of these proposals and amounts to controlling the strong and spatially dependent interactions of Rydberg atoms, which, in larger samples, lead to dephasing and prohibit the clear observation of Rabi oscillations~\cite{Deiglmayr2006, Heidemann2007, Johnson2008, Reetz-Lamour2008, Raitzsch2008, Younge2009a, Viteau2011, Schaus2012}. 
However, for small systems of up to $16$ atoms Rabi oscillations have been directly observed~\cite{Gaetan2009, Hankin2014, Barredo2014, Ebert2014}, while for larger systems indirect detection methods were required~\cite{Dudin2012b}.
In contrast, we studied isolated superatoms, locally controlling the shape on the single atom level with sub-Micrometer precision and the atom number fluctuations better than shot noise.
We detected the Rydberg excitations in-situ with single atom sensitivity and coherently manipulated collective systems with scalable size between one and $185$ individual atoms.
In this work we report on these measurements and confirm the predicted $\sqrt{N}$ enhancement of the Rabi coupling over two orders of magnitude. 
Additionally, we detect entanglement between the components of the superatom and analyze the local distribution of the Rydberg excitation within the ensemble.
Finally, we show that multiple excitations indeed lead to dephasing which, however, is reduced in the post-selected single excitation subspace.

\begin{figure*}
  \includegraphics[width=1\textwidth]{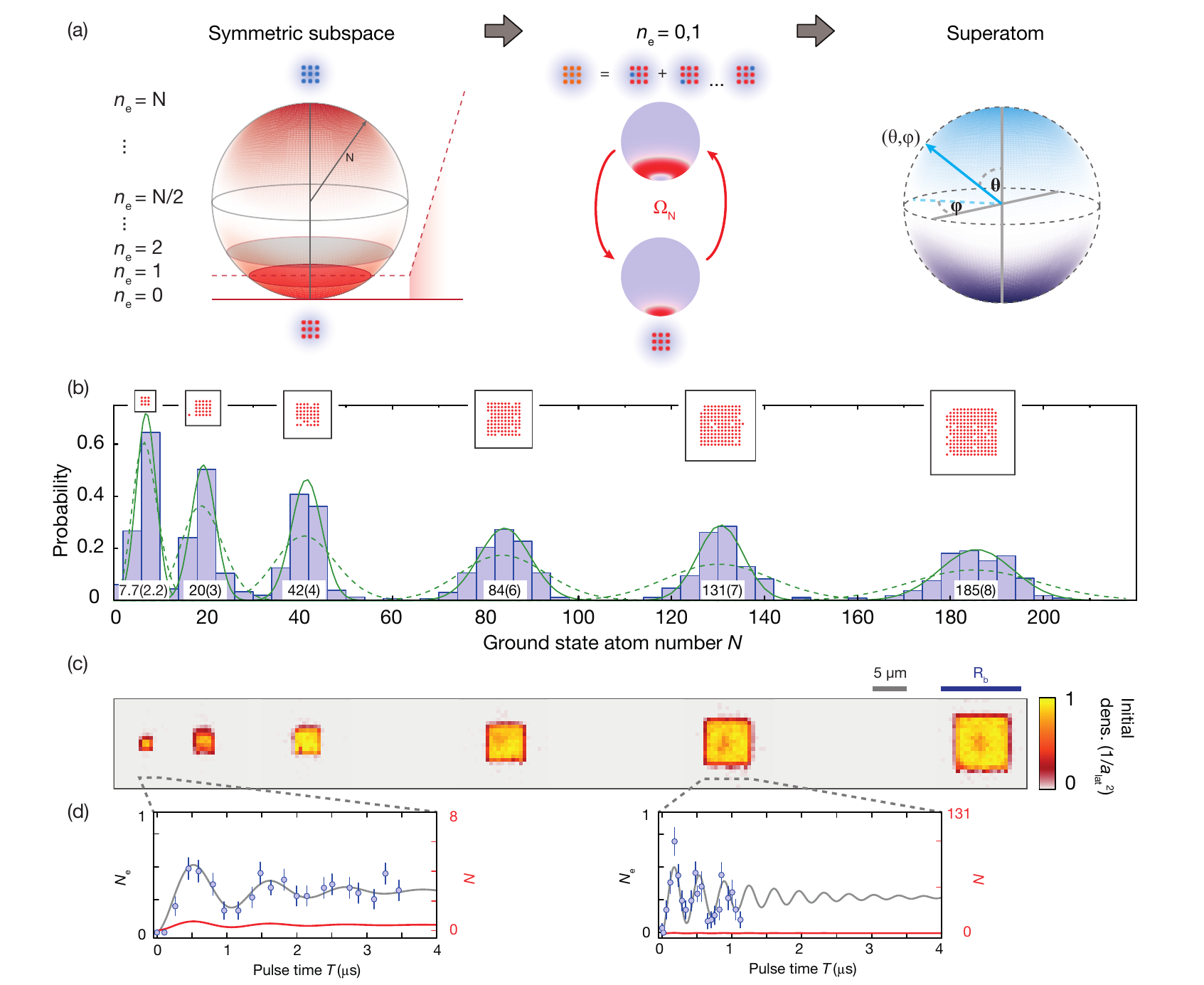}
  \caption{Superatom preparation.\label{fig:1} 
  (a) Illustration of the symmetric ground and singly excited state ($W$-state). 
  Left: $N$-atom collective Bloch sphere with excitation numbers and coupled states (south pole and $W$-state, represented by the red plane) indicated. 
  The small pictograms above and below the sphere depict the lattice system with atoms in the ground (red) and Rydberg (blue) state. 
  The Husimi distribution of these states and their enhanced coupling $\Omega_N$ is shown in the center. 
  This accessible state space defines a superatom represented by the standard Bloch sphere on the right. 
  (b) Atom number histograms of the initially prepared samples (blue bars) with Gaussian fits (solid green line). 
  The numbers give the mean and standard deviation (s.d.) for each data set.
  Measured and reconstructed occupation of lattice for exemplary initial states is depicted above the respective histograms, c.f. schematic pictograms in (a).  
  The Poissonian distribution with the same mean atom number is shown as a reference (green dashed line).
  (c) Averaged initial ground state atom distributions for the respective histograms above. 
  The size of blockade radius $R_b$ is shown by the blue bar.
  (d) Rabi oscillation data (blue points) and sinusoidal fits with exponentially decaying contrast (solid gray line) for $N$=7.7(2.2) and $N$=131(7). 
  The red line shows the same fit on an axis scaled to the number of ground state atoms $N$ (right axis). 
  All errorbars are s.e.m.
  }
\end{figure*}

\begin{figure*}
  \includegraphics{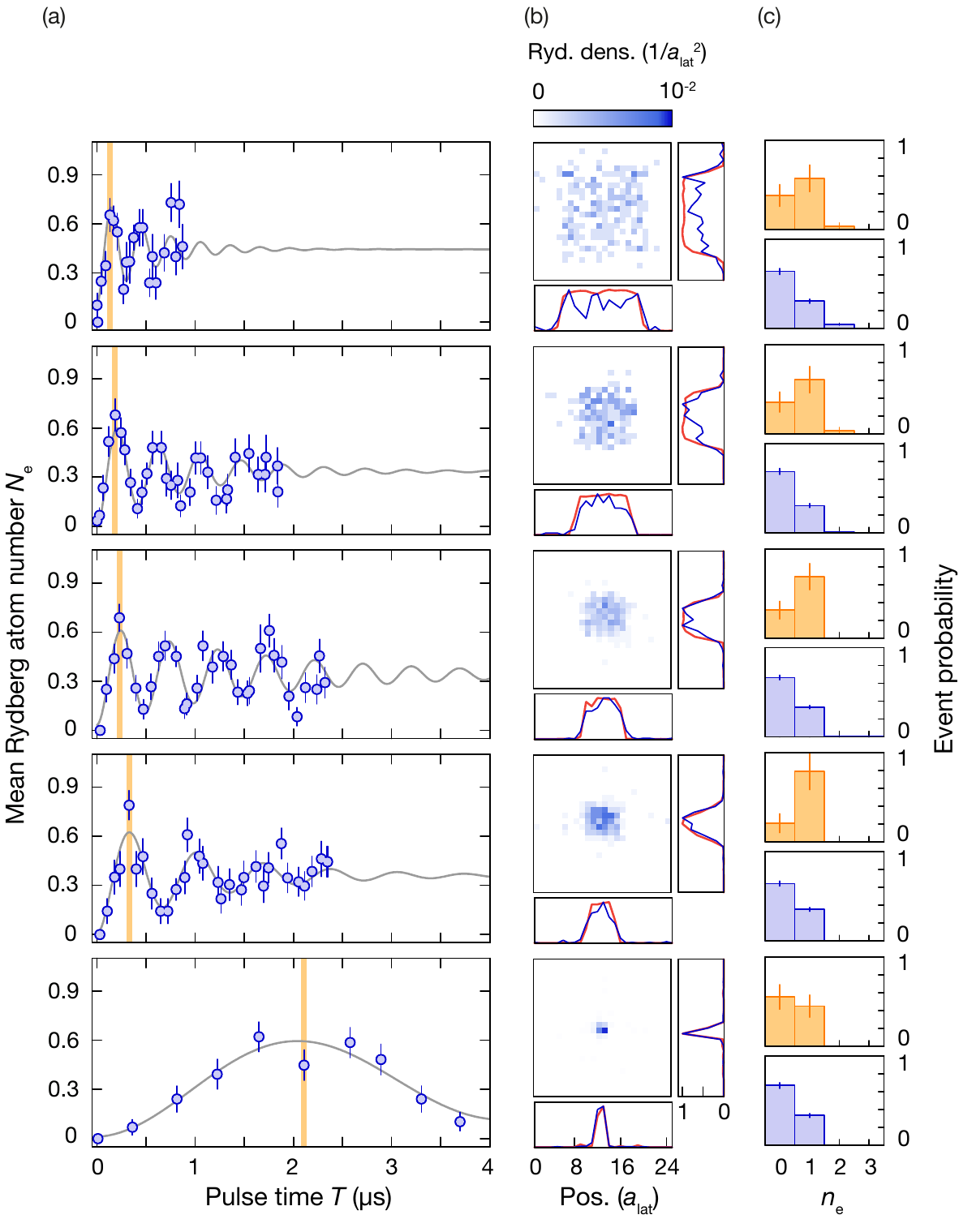}%
  \caption{Collective Rabi oscillations.\label{fig:2}
  (a) Collective Rabi oscillation data of the mean Rydberg atom number $N_e$ (blue points) for different numbers of ground state atoms $N=185(8),\ 84(6),\ 42(4),\ 20(3),\ 0.74(0.60)$ (top to bottom) with exponentially decaying sinusoidal fits (gray). 
  All errorbars are s.e.m.. 
  (b) Density of detected Rydberg atoms for the datasets in (a) with normalized vertically or horizontally averaged density (blue solid line) compared to the initial state atom distributions (red solid line). 
  (c) Histograms of the Rydberg excitation number integrated over the total oscillation (blue bars) and at the position of the first maximum (orange bars, position in (a) marked by orange solid line). 
  }
\end{figure*}
Our superatoms were formed out of an ensemble of ultracold atoms held in a two-dimensional optical lattice with unity occupation per lattice site~\cite{Schaus2012}. 
This system was then approximately uniformly coupled to a Rydberg state with coupling strength $\Omega$. 
The atoms occupied the Rydberg state only for a few Microseconds, such that their motion in the optical lattice, typically on a Millisecond time scale, could be safely neglected. 
The excited state was chosen such that for most of the experiments presented here the system size was much smaller than the dipole blockade radius $R_b$. 
This dipole blockade originates from the van-der-Waals interaction which causes an energy shift $\hbar\Delta_{\mathrm{vdW}}=\frac{C_6}{R^6}$ between Rydberg atoms separated by the distance $R$~\cite{Lukin2001}. 
The extraordinary strong interaction tunes the excitation laser out of resonance at the blockade distance $R_b=\left(\frac{C_6}{\hbar\Omega}\right)^{1/6}$ such that the system is restricted to a single Rydberg excitation. 
This single excitation is symmetrically shared among all $N$ atoms if both coupling and interaction are effectively uniform. 
Hence, the system dynamics is confined to the symmetric subspace of zero ($n_e=0$) and one ($n_e=1$) excitations, whose basis are the Fock states $\ket{0}=\ket{g_1,\dots,g_N}$ and the entangled $W$-state
$\ket{1}=\frac{1}{\sqrt{N}}\sum_{i=1}^{N}\ket{g_1,\dots,r_i,\dots,g_N}$,
where $g_i$ and $r_i$ label the i-th atom in the ground or Rydberg state.
Then, the Hamiltonian can be written in the simple form $H=\hbar\sqrt{N}\Omega/2 \left( \ket{0}\bra{1}+\ket{1}\bra{0} \right)$, where the symmetry induced collectively enhanced coupling $\sqrt{N}\Omega$ appears explicitly. 
The collective Bloch sphere (Fig.~\ref{fig:1}a) offers a convenient way of representing states within the symmetric subspace via their Husimi quasiprobability distribution~\cite{Lee1984, Arecchi1972}. 
The state $\ket{0}$ lies at the south pole of the sphere, while $\ket{1}$ corresponds to a ring-like structure. 
These many-body states form the basis for the superatom.

\begin{SCfigure*}
  \includegraphics{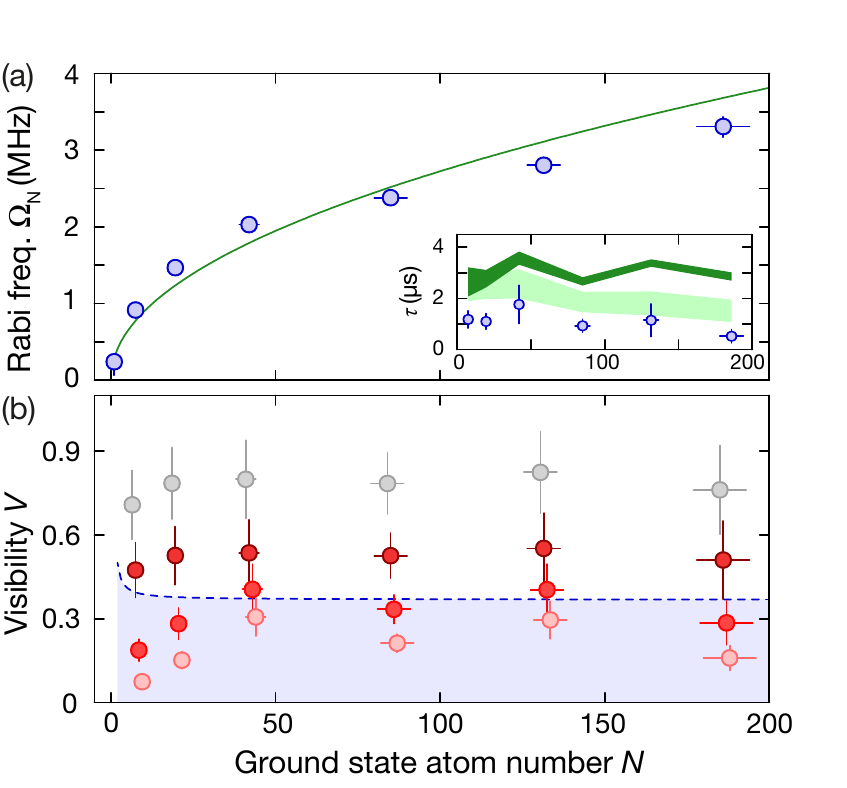}%
  \caption{Scaling of the Rabi frequency and entanglement. \label{fig:3}
  (a) Extracted values of $\Omega_N$ (blue points) versus average initial atom number $N$ for the data shown in Figs.~\ref{fig:1}d and \ref{fig:2} with a power law fit (green line). 
  The inset shows the exponential decay time of the Rabi oscillations (blue points). 
  The expected decay based on the reference sample atom number fluctuations (dark green shading) and, additionally, taking into account noise in the pulse area (light green shading) are shown for comparison. 
  (c) Extracted visibility $V$ of the collective Rabi oscillation versus atom number $N$ after one, three and five half cycles of oscillation (red data points with increasing lightness, shifted slightly horizontally for better visibility).
  The blue shaded area includes all classical states with fully separable single particle density matrices. 
  The gray points show the visibility after one half cycle corrected for the measured detection efficiency. 
  All errorbars are $1\sigma$ statistical uncertainty from the fits.
  } 
\end{SCfigure*}
For the preparation of the superatoms our experiment started with a two dimensional degenerate gas of rubidium-$87$ in the $\ket{F=2,m_F=2}$ hyperfine state, confined in a single antinode of a vertical ($z$-axis) optical lattice at $80\,E_r$. 
Here, $E_r$ is the lattice recoil energy of our square lattice with periodicity $\alat=532\,\mathrm{nm}$. 
We prepared a unity filling Mott insulator of $\sim 200-500$ atoms by adiabatically switching on two orthogonal lattices in the $x-y$ plane to $40\,E_r$. 
We then used our local addressing technique to precisely control the size and shape of the atomic sample to a square with diagonal length $D$ containing between one and $185$ atoms~\cite{Weitenberg2011, Schaus2014}. 
This ensured that the edges and the total atom number of the atomic samples were well defined, allowing us to measure total atom number fluctuations up to $4\,\mathrm{dB}$ below shot noise (Fig.~\ref{fig:1}b,c). 
The atoms were then coupled to the $68S_{1/2}, \ket{m_J=-1/2}$ Rydberg state via a two-photon scheme (red laser with wavelength $780\,\mathrm{nm}$ and blue laser with wavelength $480\,\mathrm{nm}$)~\cite{Schaus2012}. 
The excitation beams were counterpropagating perpendicular to the atomic plane ($z$-direction) with waists $w_0=44(2)\mu\mathrm{m}$ for the red and $w_0=17(5)\mu \mathrm{m}$ for the blue beam. 
The van-der-Waals coefficient of the $68S$ state is $C_6 = h \cdot 630\,\mathrm{GHz}\,\mu\mathrm{m^6}$, resulting in blockade radius of $R_b=11.7(1) \mu \mathrm{m}$ for the single particle Rabi frequency of $\Omega=2\pi \cdot 240(11)\mathrm{kHz}$. 

We detected the Rydberg atoms in-situ using an efficient ($>99.9\%$) push-out of the ground state atoms lasting $8\,\mu$s followed by a stimulated depumping of the Rydberg atoms back to the ground state. 
The remaining atoms were then imaged using in-situ fluorescence detection with a position resolution of approximately $\pm 1$ lattice site~\cite{Schaus2012}. 
From our data we inferred an overall efficiency of $\eta=0.67(5)$ for the spatially resolved detection of a single Rydberg atom.
The spatial control over the sample allowed for microscopic control of the superatom size. 
As long as we ensured $R_b \gg D$ (up to and including $131$ atoms) we observed coherent enhanced Rabi oscillations between the zero and one excitation subspaces. Here, the ratio of the amplitude of the Rabi oscillations to the total atom number scales as $1/N$ as opposed to being constant for independently oscillating particles. 
In Fig. ~\ref{fig:1}d we illustrate this scaling for two exemplary cases of $8$ and $131$ atoms.


In order to characterize the prepared superatoms microscopically, we measured the spatial distribution of the observed Rydberg atom and the excitation statistics during the Rabi oscillation (Fig.~\ref{fig:2}). 
For different sample sizes between one and $185$ atoms we drove Rabi oscillations by illuminating the sample with the coupling lasers for varying duration $T$. 
For each $T$ we repeated the experiment $25-30$ times and extracted the mean Rydberg number $N_e$ (Fig.~\ref{fig:2}a). 
The dramatic acceleration of the Rabi oscillation with $N$ is clearly visible in the data. 
Additionally, we compare the spatial distribution of the Rydberg atoms (integrated over all $T$) to the initial distribution of ground state atoms. 
Within statistical uncertainty we find a flat distribution consistent with the uniform coupling assumption (Fig.~\ref{fig:2}b). 
We experimentally confirm the picture of a fully dipole blockaded sample by extracting the histogram of the Rydberg excitation numbers $n_e$ both integrated over the whole observation time $T$ and as well at the $\pi$-pulse time $T_\pi$. 
For sample sizes up to $131$ atoms, the probability of measuring doubly excited states with two detected Rydberg atoms was below $1\%$. We obtained typically $1-4$ images with two excitations per $500-800$ shots. 
For the largest sample used in our experiments, the blockade starts to break down and the probability increased to $4.8(1.0)\%$ ($27$ events per $564$ shots). 
None of the data shown here was corrected for the detection efficiency and the measured excitation number after $T_\pi$ is consistent with $N_e=1$ when taking it into account.

One striking signature of the superatom is its symmetry-enhanced coupling to the radiation field. 
We extracted the oscillation frequency $\Omega_N$, the decay time $\tau$ and a global offset $A$ from the data shown in Fig.~\ref{fig:1}d and \ref{fig:2}a via a fit to  
$N_{e}=\eta\cdot\left(A-e^{-t/\tau}\cos\left(\Omega_{N}t\right)/2\right)$. 
Indeed, we confirm the expected scaling $\Omega_N \propto \sqrt{N}$ over two orders of magnitude (Fig.~\ref{fig:3}a). 
A power law fit of the form $\Omega_N=\Omega\cdot N^{\alpha}$ yields an exponent of $\alpha=0.48 (10)$. 
Deviations towards higher Rabi frequencies for small $N$ might be due to a residual detuning of the coupling lasers that we calibrated via spectroscopy on a dilute atomic cloud with an uncertainty of $\pm200~\mathrm{kHz}$. 
A systematic lower Rabi frequency at large $N$ can be caused by a residual inhomogeneity of the laser coupling (up to $10\%$) due to its Gaussian intensity profile. 
Also, the observed onset of a breakdown of the blockade for our largest prepared samples results in a deviation from the $\sqrt{N}$-scaling. 
The latter effect is additionally visible in the extracted steady state mean Rydberg atom number $\eta A$. 
For all but the largest sample size we find $A=0.51(2)$, consistent with the expected value of $0.5$ . 
For $N=185$ it increases significantly to $A=0.65(12)$.
To answer the question whether the collective speedup can be exploited for quantum operations or whether decoherence effects dominate, we analyzed the quality factor of the Rabi oscillations, which is given by the product of the decay time $\tau$ of the measured oscillations and the Rabi frequency $\Omega_N$.
Indeed, we find a peaking quality factor for $N=131$ due to the increasing oscillation frequency but constant decay time $\tau\approx 1\,\mu$s in the fully blockaded regime (Fig.~\ref{fig:3}a inset).
Among the limiting factors for the coherence time are residual atom number and coupling power fluctuations ($8(2)\%$). 
However, these alone cannot explain the observed decay~(Fig.~\ref{fig:3}a inset). 
For small atom numbers additional decoherence might be due to phase noise and slow frequency drifts of the lasers, while the inhomogeneity in the Rabi coupling becomes significant at larger $N$.


Next to the collective enhancement of the optical coupling, the structure of the excited state itself bears the marks of the strong particle correlations. 
The unambiguous proof that the excited state of the superatom is indeed the $N$ particle entangled $W$-state would require full state tomography~\cite{Haffner2005}, which is not feasible in our setup. 
However, we will show that the experimental observations are incompatible with the expectations for a fully separable state. 
To this end, we employ the entanglement witness developed in Ref.~\cite{Haas2014} which compares the overlap of the $W$-state and the prepared state to the maximally possible overlap with any separable state. 
In order to extract the $W$-state overlap from the measured oscillation we model the dynamics by a sum of density matrices, $\rho_\mathrm{meas}=\alpha(t)\rho_c(t)+\beta(t)\rho_m$, where the probabilities $\alpha(t)$ and $\beta(t)$ interpolate between the coherently oscillating part described by $\rho_c(t)$ and a mixed density matrix $\rho_m$ showing no dynamics. 
Under the experimentally verified assumption that maximally one excitation was present in the system, $\rho_c(t)$ describes the subspace spanned by the ground and the $W$-state, as population of all other singly excited states leads to dephasing of the oscillations of the uniformly coupled system.
Hence, $\alpha(t)$ can be identified with the overlap of the prepared state with the $W$-state and it is directly measured by the visibility $V(t)$ of the Rabi oscillations.
In Fig. ~\ref{fig:3}b we show that the extracted $V$ at the oscillation maximum after the first half Rabi cycle is consistently above the threshold for entanglement, even without correcting for the detection efficiency $\eta$, while it decays into the classically allowed region for longer times.
If the detection efficiency $\eta$ is taken into account, we obtain a lower bound for the average $W$-state preparation fidelity of $0.77(20)$ at the first maximum.

\begin{figure}
  \includegraphics{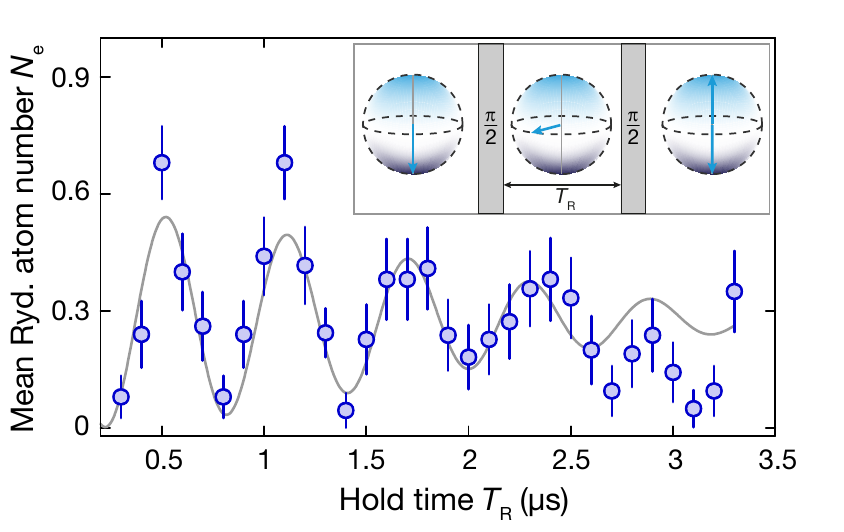}%
  \caption{Superatom Ramsey spectroscopy. \label{fig:4} 
  Evolution of the mean Rydberg atom number $N_e$ (blue points) versus hold time $T_R$ between the two $\pi/2$-pulses (schematic in the inset) and sinusoidal fit with Gaussian decay (gray line) for an initial sample size of $N=38(3)$ atoms. 
  All errorbars are s.e.m.
  }
\end{figure}

To investigate the coherence of the collective Qubit further, we used a $N=38(3)$ atom ensemble for Ramsey interferometry~\cite{Ebert2015}. 
First, we prepared a coherent superposition of $\ket{0}$ and $\ket{1}$ by a $\pi/2$-pulse of length $T=\pi/(2\Omega\sqrt{N})$. 
After a variable hold time $T_{R}$, a second $\pi/2$-pulse was applied and the mean Rydberg atom number $N_e$ was measured (Fig.~\ref{fig:4}).
Due to the ac-Stark shift created by the red $780\,\mathrm{nm}$ laser during the excitation pulse, the calibrated transition frequency differs from the bare ground to Rydberg state transition frequency, which defines the reference for the Ramsey interferometer.

Therefore, the observed phase accumulation rate is given by this ac-Stark shift and agrees well with an independent calibration of the latter via microwave spectroscopy.
The extracted decay time of the Ramsey fringe of $\tau_{R}=2.2(4)\mu \mathrm{s}$ exceeds the damping $\tau$ of the Rabi oscillations, indicating that there laser power fluctuations and coupling inhomogeneities are likely the main source of decoherence. 
We also performed Ramsey interferometry for different $N$ and find within experimental uncertainty $\tau_{R}$ approximately independent of $N$.

\begin{SCfigure*}
  \includegraphics{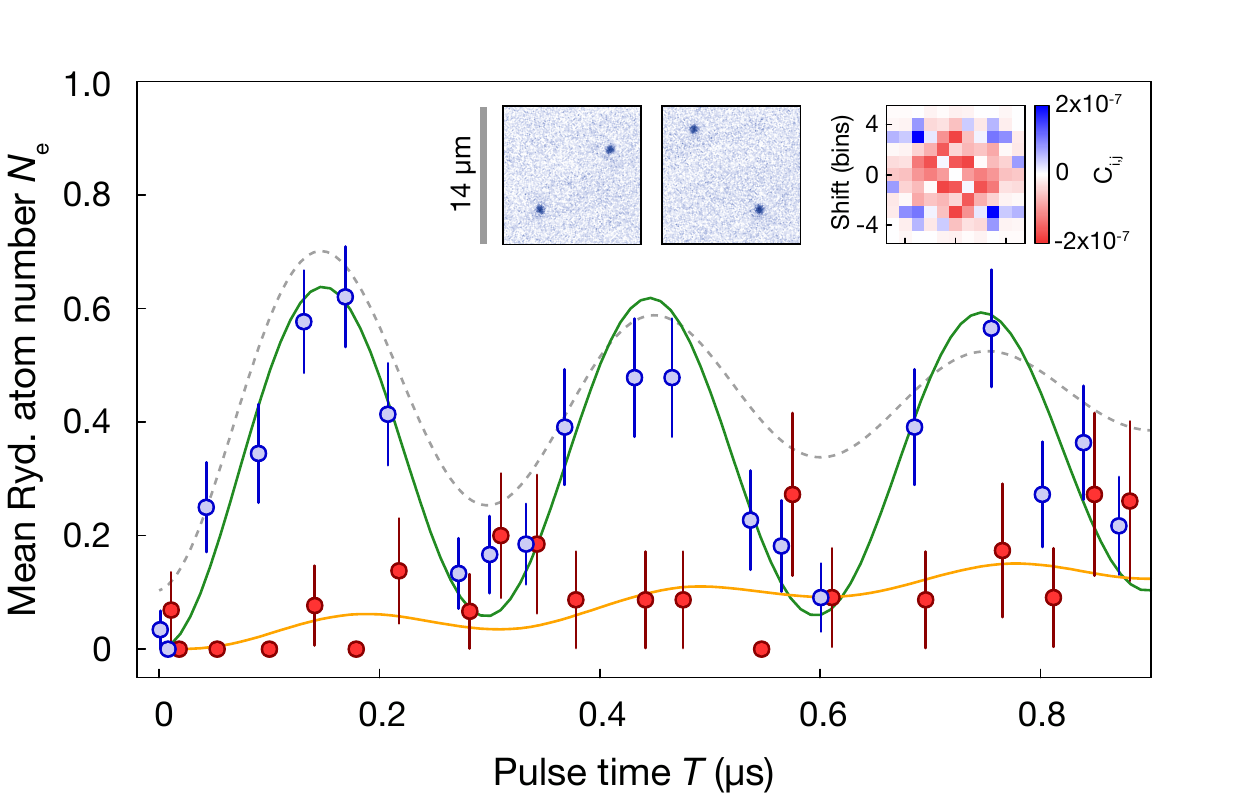}%
  \caption{Breakdown of the blockade. \label{fig:5}
  Measurement of the collective Rabi oscillation in sample with $N=185(8)$ atoms. 
  (a) The contribution of states with $n_e=1$ (blue points) and $n_e=2$ (red points, shifted slightly horizontally for better visibility) agree with the theoretical calculation (green and orange solid lines, scaled by extracted detection efficiency $\eta$). 
  For comparison we show the fit (gray line) to the mean (c.f. Fig.~\ref{fig:2}), which shows twice faster dephasing than the $n_e=1$ subspace alone.
   The inset shows two exemplary single shots with $n_e=2$ (field of view $14\times14\,\mu\mathrm{m}$, indicated by gray bar) and the two-dimensional pair-correlation function $C_{i,j}$ of all $n_e=2$ events. 
  Color scale: red (anti-correlation) to blue (correlation). 
  Data binned $4\times 4$ sites. 
  All errorbars are s.e.m. 
  }
\end{SCfigure*}
When increasing the sample to a size where the maximal distance between two atoms $D$ approaches the blockade radius, the isolated superatom-picture is expected to break down~\cite{Weber2015}. 
The gap to doubly excited states with two atoms pinned to the diagonal corners of the prepared square shaped density distribution is smallest, such that these are the first doubly excited states populated. 
We discussed already several indications of this blockade breakdown for the $N=185$ atom sample. 
Here, the maximum separation of two atoms was $D=9.8(7)\,\mu\mathrm{m}$, close to the blockade radius $R_b=11.7\,\mu \mathrm{m}$. 
In Fig. ~\ref{fig:5} we study the effects of the doubly excited states on the decay of the Rabi oscillation for this setting by post-selecting the acquired data to single and double Rydberg events. 
The decay time of the singly excited component is two times larger compared to the full sample and agrees with the prediction of a numerical calculation of the system in a reduced Hilbert space~\cite{Schaus2012}. 
This shows that the observed decay is significantly influenced by the dephasing due to double excitation. 
We observe a slow increase of the doubly excited fraction (also in agreement with theory) that is consistent with the picture of two interacting excitations~\cite{Stanojevic2009}. Their interaction energy corresponds to the energy shift $\Delta_{\mathrm{vdW}}$ at the distance $D$, resulting in a detuned optical coupling. 
At the same time the collective enhancement of the coupling to this state is only $\sqrt{2}$ reflecting the two possible orientations along the diagonals. 
The resulting time scale $\pi/\sqrt{2\Omega^2 + \Delta_{\mathrm{vdW}}^2}$ matches roughly the observed slow rise of the doubly excited states, however, explaining their probability quantitatively requires a more complex model including also atoms close to the corners of the square.   
Spatial correlation measurements confirm the localization of the doubly excited events at the diagonal corners (Fig.~\ref{fig:5} inset). 
Low statistics requires here the binning of $4\times 4$ lattice sites for the evaluation of the two point correlation $C_{i,j}= \langle\langle P_{(x,y)}P_{(x+i,y+j)}\rangle-\langle P_{(x,y)}\rangle \langle P_{(x+i,y+j)}\rangle\rangle_{x,y}$. 
Here, $P_{(i,j)}$ is the probability of finding a Rydberg excitation in bin $(i,j)$ and $\langle \cdot\rangle_{x,y}$ and $\langle \cdot\rangle$ denote the spatial and ensemble averages. 

In conclusion, we demonstrated coherent control and two-orders of magnitude scalability of Rydberg-superatoms. 
Using in-situ microscopical detection of the Rydberg atoms we confirmed the superatom picture and proved presence of entanglement in the involved singly excited many-body states. 
We also demonstrated that the collectively enhanced coupling can be harnessed to increase the fidelity of collective Qubit rotations under realistic experimental conditions.
The experiments confirmed that coupling to many-body states with larger Rydberg occupation leads to interaction induced dephasing, strongly supporting the coherent description of our previous experiment on short time scales~\cite{Schaus2012, Schaus2014}.
Our results pave the way towards the controlled step-wise preparation of higher Dicke states~\cite{Lukin2001}, which have been proposed for metrology at the Heisenberg limit~\cite{Holland1993}, and they promise to shed light on macroscopic entangled quantum systems~\cite{Opatrny2012}.

\begin{acknowledgments}
We thank Marc Cheneau and Takeshi Fukuhara for valuable discussions and Jae-yoon Choi for proofreading the manuscript. T.M. thanks the International Institute of Physics (Brazil) for the hospitality during the completion of the paper. We acknowledge funding by MPG, EU (UQUAM, SIQS) and the K\"orber Foundation.
\end{acknowledgments}

\bibliography{superatom_arxiv}

\end{document}